\documentclass[11pt,onecolumn]{article}

\usepackage{epsfig}
\usepackage{amsmath}
\usepackage{amssymb}
\usepackage{indentfirst}
\usepackage{cite}
\usepackage{bm}
\usepackage{graphicx}           
\usepackage{psfrag}             
\usepackage{empheq}
\usepackage{enumerate}

\usepackage{wrapfig,xargs,mathtools,mathrsfs,microtype,titlesec,url,blindtext,nicefrac,comment,algorithm}
\usepackage[font=small]{caption}
\usepackage{color}
\usepackage[dvipsnames]{xcolor}

\newcommand{\HB}[1]{{\color{blue}#1}}

\newcommand{\diag}{\textnormal{diag}}

\usepackage{color}
\definecolor{gray}{RGB}{128,128,128}

\usepackage{amsthm}

\usepackage{tikz}
\usetikzlibrary{arrows}

\newenvironment{proof}[1][Proof]%
  {\smallskip\par\noindent\textbf{#1\,:\ }}%
  {\hspace*{\fill} \rule{6pt}{6pt}\smallskip}
\newenvironment{proof*}[1][Proof]%
  {\medskip\par\noindent\textbf{#1\,:\ }}%

\newtheorem{assumption}{Assumption}

\usepackage{hyperref}

\title{\bf Hierarchical Control of Multi-Agent Systems using Online Reinforcement Learning}

\author{He Bai,~Jemin George~and~Aranya Chakrabortty
\thanks{H.~Bai is with Oklahoma State University, Stillwater, OK 74078, USA.
{\tt\small he.bai@okstate.edu}}%
\thanks{J.~George is with the U.S. Army Research Laboratory, Adelphi, MD 20783, USA.
{\tt\small jemin.george.civ@mail.mil}}%
\thanks{A. Chakrabortty is with  North Carolina State University, Raleigh, NC 27695, USA.
{\tt\small achakra2@ncsu.edu}}%
}

\allowdisplaybreaks[4]

\begin{document}

\maketitle
\thispagestyle{empty}
\pagestyle{empty}

\begin{abstract}
We propose a new reinforcement learning based approach to designing hierarchical linear quadratic regulator (LQR) controllers for heterogeneous linear multi-agent systems with unknown state-space models and separated control objectives. The separation arises from grouping the agents into multiple non-overlapping groups, and defining the control goal as two distinct objectives. The first objective aims to minimize a group-wise block-decentralized LQR function that models group-level mission. The second objective, on the other hand, tries to minimize an LQR function between the average states (centroids) of the groups. Exploiting this separation, we redefine the weighting matrices of the LQR functions in a way that they allow us to decouple their respective algebraic Riccati equations. Thereafter, we develop a reinforcement learning strategy that uses online measurements of the agent states and the average states to learn the respective controllers based on the approximate Riccati equations. Since the first controller is block-decentralized and, therefore, can be learned in parallel, while the second controller is reduced-dimensional due to averaging, the overall design enjoys a significantly reduced learning time compared to centralized reinforcement learning. 

\end{abstract}

\section{Introduction}
Conventional reinforcement learning (RL) based control of high-dimensional LTI systems with unknown state-space models using algorithms such as actor-critic methods \cite{lewis}, Q-learning \cite{vam2}, and adaptive dynamic programming (ADP) \cite{jiang} usually have long learning times. This is because the initialization of these learning algorithms involves a least squares estimation step that requires RL to wait until a minimum amount of time for gathering sufficient amount of state and input data so that the appropriate data matrices can be guaranteed to have full rank. Larger the size of the plant, more is this waiting time. 

In this paper we show that reduction in learning time is possible in scenarios where the control objective can be decomposed into a hierarchy of sub-objectives. We consider a large number of self-actuated agents with decoupled open-loop heterogeneous LTI dynamics. The state and input matrices of each agent are assumed to be unknown, while all states and control inputs are assumed to be measured. The agents are assumed to be divided into multiple non-overlapping groups that may arise from various factors such as their geographical proximity, the nature of their mission, or the physical characteristics of the agents. The grouping is imposed only to make the separation in the control objectives well-defined; it does not imply any kind of time-scale separation or redundancy in controllability as in \cite{sayakcdc}.  The goal is to learn two distinct sets of LQR controllers. The first controller is a {\it local} LQR controller for each individual group that employs feedback of only the agent states belonging to that group. The overall local control gain matrix is thus block-diagonal in structure. The second controller, on the other hand, is a {\it global} LQR controller that is meant to control the relative motion between the centroids of each group. It is, therefore, a reduced-dimensional controller that employs feedback from only the average states of each group.



A model-based version of this hierarchical LQR control for homogeneous LTI models was recently reported in \cite{shinji}, followed by other optimization based designs in \cite{hara,imura,javadnew}. Motivated by the technique presented in \cite{shinji}, we first redefine the input weighting matrices of the LQR functions in a way that they allow us to decouple their respective algebraic Riccati equations (AREs). Our approach is  different from \cite{shinji} where instead the state weighting matrix was redefined, leading to a different set of Riccati equations. Thereafter, we develop a reinforcement learning strategy using ADP  to learn the two respective controllers based on the redefined approximate Riccati equations. Since the first controller is block-decentralized and, therefore, can be learned in parallel, while the second controller is reduced-dimensional due to averaging, the overall design enjoys a significantly reduced learning time compared to centralized reinforcement learning. We illustrate the effectiveness of the design and also highlight its drawbacks on sub-optimality using an example from hierarchical formation control.

The rest of the paper is organized as follows. Section II formulates the hierarchical LQR control problem, and provides its model-based solution using approximations in the AREs. Section III develops a variant of ADP that learns the hiearchical controllers using state and input measurements. Section IV shows the applicability of this design for multi-agent formation control, illustrated with numerical simulations. Section V concludes the paper.

\section{Problem Formulation}
Consider a multi-agent network consisting of $N>0$ groups. For $j=1,\dots,N$, the $j^{th}$ group contains $p_j$ number of agents, with any agent $i$ satisfying the dynamics
\begin{equation}
    \dot{{x}}_i = G_i {x}_i + H_i {u}_i, \label{plant}
\end{equation}
 where ${x}_i \in \mathbb{R}^n$ is the state, and ${u}_i \in \mathbb{R}^m$ is the control input of the agent, for all $i = 1,\ldots,p$, $p=\sum_j^N{p_j}$. The matrices $G_i$ and $H_i$ are unknown, although their dimensions are known. The agents are assumed to be initially uncoupled from each other. Let  $\mathbf{x}_j$  and $\mathbf{u}_j$ represent the vector of all states and control inputs in group $j$. The group-level dynamics are written as
 \begin{equation}
     \mathbf{\dot x}_j={A}_j\mathbf{x}_j+{B}_j\mathbf{u}_j,\label{tr}
\end{equation}
where $A_j$ and $B_j$ are block-diagonal concatenations of $G_i$ and $H_i$, respectively, for all agents $i$ belonging to group $j$. Denoting $\mathbf{x} = \begin{bmatrix} \mathbf{x}^\top_1 & \ldots & \mathbf{x}^\top_N \end{bmatrix}^\top \in \mathbb R^{pn}$, $\mathbf{u} = \begin{bmatrix} \mathbf{u}^\top_1 & \ldots & \mathbf{u}^\top_N \end{bmatrix}^\top \in \mathbb R^{pm}$, the  network model becomes
\begin{equation}
    \dot{\mathbf{x}} = \mathcal{A} \mathbf{x} + \mathcal{B} \mathbf{u}, \label{tot}
\end{equation}
where $\mathcal{A} \in \mathbb{R}^{np\times np}$ and $\mathcal{B} \in \mathbb{R}^{np\times mp}$ are block-diagonal matrices consisting of $A_j$'s and $B_j$'s, respectively.

Let the control objective be to design a state-feedback controller $\mathbf{u}=-K\mathbf{x}$ to minimize the cost
\begin{equation}
    J = \int_0^\infty \mathbf{x}^\top Q\mathbf{x} + \mathbf{u}^\top R\mathbf{u} \,\, dt, \label{J}
\end{equation}
where $Q \geq 0$ and $R > 0$ are performance matrices of appropriate dimensions, constrained to \eqref{tot}.

\begin{assumption}
The communication topology among the centroids of the $N$ groups is an undirected network with a Laplacian matrix $L \in \mathbb R^{N \times N}$.
\end{assumption}

\begin{assumption}
Performance matrices $Q$ and $R$ are given as
\begin{align}
    R &= \diag\{ R_1, \ldots, R_N\},\,  Q = \bar{Q} + L_w \odot \tilde{Q}\label{q},\\
    \bar{Q} &= \diag\{\bar{Q}_1, \ldots, \bar{Q}_N\},
\end{align}
where $R_j \in \mathbb{R}^{mp_j\times mp_j} > 0$, $\bar{Q}_j \in \mathbb{R}^{np_j\times np_j} \geq 0$ for all $j = 1,\ldots,N$, $L_w$ is a weighted Laplacian matrix which has the same structure as $L$, and
\begin{eqnarray}
\tilde{Q} =  \begin{bmatrix} \frac{1}{p_1^2}\mathbf{1}_{p_1}\mathbf{1}_{p_1}^\top  &  \ldots & \frac{1}{p_1p_N}\mathbf{1}_{p_1}\mathbf{1}_{p_N}^\top \\
\frac{1}{p_2p_1}\mathbf{1}_{p_2}\mathbf{1}_{p_1}^\top &  \ldots & \frac{1}{p_2p_N}\mathbf{1}_{p_2}\mathbf{1}_{p_N}^\top \\
\vdots & \ddots & \vdots \\
\frac{1}{p_NP_1}\mathbf{1}_{p_N}\mathbf{1}_{p_1}^\top &  \ldots & \frac{1}{p_N^2}\mathbf{1}_{p_N}\mathbf{1}_{p_N}^\top \end{bmatrix} \otimes I_n.  \label{qtilde}
\end{eqnarray}
It can be seen that $\tilde{Q}\in \mathbb{R}^{pn\times pn} \geq 0$. Here, $\otimes$ is the Kronecker product and $\odot$ is the Khatri$-$Rao product.
\end{assumption}

The two components of the matrix $Q$ in \eqref{q} represent the separation in the control objective. The block-diagonal component $\bar{Q}$ represents group-level local objective, such as maintaining a desired formation for each group in the multi-agent network. 
The second component, on the other hand, represents a global objective that is meant to coordinate a set of {\it compressed} state vectors chosen from each group. In this case, as indicated in \eqref{qtilde} we assume the compressed state to be simply the centroid, i.e., the average of the respective group states. More general definitions of this compressed state is possible, but we stick to this assumption for simplicity. Denote the centroid state of the $j^{th}$ group by $\mathbf{x}_{av,j} \in \mathbb R^{n}$, which is the average of the state vectors of all the agents in that group. Let the quadratic objective for the $j^{th}$ group  be
\begin{align} \label{jav}
    J_{av,j} &= \sum_{\ell \in \mathcal N_j} (\mathbf{x}_{av,j}-\mathbf{x}_{av,\ell})^\top \mathcal Q_{j\ell}(\mathbf{x}_{av,j}-\mathbf{x}_{av,\ell}),
\end{align}
where $\mathcal N_j$ is the neighbor set of the $j^{th}$ centroid following the structure of the Laplacian matrix $L$, and $\mathcal Q_{j\ell} \geq 0$ is a given $n\times n$ design matrix.

Since the network is assumed to be undirected, the Laplacian matrix can be written as $L = DD^\top$, where $D$ is the incidence matrix. Define the weighted Laplacian as
\begin{equation}
    L_w = (D\otimes I_n)\mathcal{Q}(D^\top\otimes I_n),
\end{equation}
where $\mathcal{Q}$ is a block-diagonal matrix with $\mathcal Q_{j\ell}$'s as diagonal entries. Equation  \eqref{jav} for the entire network can be written as
\begin{equation} \label{jav2}
    J_{av} = \sum_{j}^N J_{av,j} = \mathbf{x}_{av}^\top \,L_w\,\mathbf{x}_{av},
\end{equation}
where $\mathbf{x}_{av} = \begin{bmatrix} \mathbf{x}^\top_{av,1} & \ldots & \mathbf{x}^\top_{av,N} \end{bmatrix}^\top \in \mathbb R^{nN}$. Also, since $$\mathbf{x}_{av} := \left(\underbrace{\diag\left\{\frac{\mathbf{1}_{p_1}^\top}{p_1},\,\frac{\mathbf{1}_{p_2}^\top}{p_2},\,\dots, \frac{\mathbf{1}_{p_N}^\top}{p_N}\right\}}_M\otimes I_n \right) \mathbf{x}$$, \eqref{jav2} can be further written as
\begin{eqnarray}
    J_{av} &= &\mathbf{x}^\top \,\left(M \otimes I_n \right)^\top\,L_w\,\left(M \otimes I_n \right) \mathbf{x}, \\
    &=& \mathbf{x}^\top \,  L_w \odot \tilde{Q} \,     \mathbf{x}
    \label{jav3}
\end{eqnarray}
which justifies the definition of $\tilde{Q}=M^\top M$ in \eqref{qtilde}.

\begin{assumption}
$\left(\mathcal{A}, \mathcal{B}\right)$ is controllable and $\left(Q^{1/2}, \mathcal{A}\right)$ is observable.
\end{assumption}

The optimal control input for minimizing \eqref{J} is
\begin{equation}
    \mathbf{u} = -K^* \mathbf{x} = -R^{-1}\mathcal{B}^\top P^*,
\end{equation}
where $P^*\in \mathbb{R}^{np \times np}$ is the unique positive definite solution of the following Riccati equation:
\begin{equation}
    {P^*}\mathcal{A}+\mathcal{A}^\top{P^*}+Q-{P^*}\mathcal{B}{R}^{-1}\mathcal{B}^\top{P^*}=0. \label{ric}
\end{equation}
As $(\mathcal{A},\,\mathcal{B})$ are unknown, \eqref{ric} cannot be solved directly. Instead it can be solved via RL using measured values of $\mathbf{x}$ and $\mathbf{u}$. One may disregard the separation property in \eqref{q}, and solve for $P^*$ from \eqref{ric} using the centralized RL algorithm in \cite{lewis}, but the drawback in that case will be a long learning time owing to the large dimension of $P^*$. The benefit will be that $P^*$ is optimal. Our approach, in contrast, is to make use of the separation property in \eqref{q} to learn and implement an RL controller using two separate and parallel components, thereby reducing the learning time. In fact, as will be shown next, the learning phase in this case reduces to learning only the individual group-level local controllers. Once learned, these local controllers can be used to compute the global component of $\mathbf{u}$ through a simple matrix product. The drawback is that the learned $\mathbf{u}$ is no longer optimal. We next describe the construction of this sub-optimal control input using an approximation for $R$.

\subsection{Approximate Control}

Define $\mathcal{P} = \diag\{P_1,\ldots,P_N\}$, where $P_j \in \mathbb R^{np_j \times np_j}$ are symmetric positive-definite matrices. Define $P_{Aj} = P_j{A}_j$ and $P_{Bj} = P_j{B}_jR_j^{-1}{B}_j^\top P_j$ for $j=1,\ldots,N$. Following \cite{shinji}, we also define
\begin{equation}
    \mathcal{R}^{-1} = {R}^{-1} + \tilde{R}, \label{rtil}
\end{equation}
where the expression for $\tilde{R}$ will be derived shortly. Then,
\begin{align}
    &\mathcal{P}\mathcal{A}+\mathcal{A}^\top\mathcal{P}+Q-\mathcal{P}\mathcal{B}\mathcal{R}^{-1}\mathcal{B}^\top\mathcal{P} \nonumber \\
    &= \mathcal{P}\mathcal{A}+\mathcal{A}^\top\mathcal{P} + \bar{Q} -
    \mathcal{P}\mathcal{B}{R}^{-1}\mathcal{B}^\top\mathcal{P} + L_w \odot \tilde{Q} - \mathcal{P}\mathcal{B}\tilde{R}\mathcal{B}^\top\mathcal{P} \nonumber  \\
    &= \diag\{P_{A1} + P_{A1}^\top + \bar{Q}_1 - P_{B1},\ldots,P_{A_N} + P_{A_N}^\top + \bar{Q}_N \nonumber \\ & - P_{B_N}\} + L_w \odot \tilde{Q} - \mathcal{P}\mathcal{B}\tilde{R}\mathcal{B}^\top\mathcal{P}. \label{str}
\end{align}

Compared to the approximation suggested in \cite{shinji}, we propose to fix $Q$ and instead adjust $R$ to account for the coupled terms in $Q$. Adjusting $R$ is more amenable for this design since perturbing $Q$ will severely degrade the system performance while adjusting $R$ will only increase (or decrease) the control demand. Furthermore, considering the structure of the RHS of \eqref{str}, it is easier to choose $R$ to cancel out the coupling term $L_w \odot \tilde{Q}$ than choosing $Q$.  Thus, if $\tilde{R}$ is selected so that
\begin{equation}
    \mathcal{P}\mathcal{B}\tilde{R}\mathcal{B}^\top\mathcal{P} = L_w \odot \tilde{Q}, \label{r}
\end{equation}
then each individual matrix $P_j$ satisfies
\begin{equation}
    P_jA_j + A_j^\top P_j + \bar{Q}_j - P_jB_jR_j^{-1}B_j^\top P_j = 0,  \label{locric}
\end{equation}
for $j=1,\dots,N$. The control gain follows as
\begin{equation}
    K = \mathcal{R}^{-1}\mathcal{B}^\top\mathcal{P}= \underbrace{R^{-1}\mathcal{B}^\top\mathcal{P}}_{local}+\underbrace{\tilde{R}\mathcal{B}^\top\mathcal{P}}_{global}.\label{eq:K}
\end{equation}
Note that the global component does not need to be learned. Once $\mathcal P$ is learned from \eqref{locric} the global controller can simply be computed using $\mathcal P$, $\tilde{R}$ and $\mathcal{B}$. As $B^\top P$ is block diagonal, the structure in $\tilde R$ dictates the structure of the global control.

The problem, however, lies in the fact that it may be difficult to find a $\tilde{R}$ which satisfies \eqref{r}. If $\mathcal B$ is a square full rank matrix (i.e., each agent is a fully actuated system) then $\tilde{R}$ follows in a straightforward way by computing the inverse of the square matrices $\mathcal P \mathcal B$ and $\mathcal B^\top \mathcal P$. Otherwise, one has to compute a least square estimate for $\tilde{R}$ as
\begin{equation}
    \tilde{R}^*
    =  \left( \mathcal{B}^\top\mathcal{P} \mathcal{P}\mathcal{B} \right)^{-1} \mathcal{B}^\top\mathcal{P} \left(L_w \odot \tilde{Q}\right) \mathcal{P}\mathcal{B} \left( \mathcal{B}^\top\mathcal{P} \mathcal{P}\mathcal{B} \right)^{-1}. \label{ls}
\end{equation}
Since $\mathcal{P}\mathcal{B}$ is block-diagonal, we can write it as
\begin{equation}
\mathcal{P}\mathcal{B}=I_N\odot \mbox{diag}\{P_1B_1,\cdots,P_NB_N\}.
\end{equation}
The matrix $\left( \mathcal{B}^\top\mathcal{P} \mathcal{P}\mathcal{B} \right)^{-1}$ can be written in a similar fashion. In that case, it follows that
\begin{align}
    &\tilde{R}^*
    =\left( \mathcal{B}^\top\mathcal{P} \mathcal{P}\mathcal{B} \right)^{-1} \mathcal{B}^\top\mathcal{P} \left(L_w \odot \tilde{Q}\right) \mathcal{P}\mathcal{B} \left( \mathcal{B}^\top\mathcal{P} \mathcal{P}\mathcal{B} \right)^{-1}\nonumber\\
    &=(I_N\odot\mbox{diag}\{(B_i^\top P_iP_iB_i)^{-1}\})(I_N\odot\mbox{diag}\{B_i^\top P_i\}) \nonumber \\ &\left(L_w \odot \tilde{Q}\right)(I_N\odot\mbox{diag}\{P_iB_i\})(I_N\odot\mbox{diag}\{(B_i^\top P_iP_iB_i)^{-1}\}) \nonumber \\
    &=L_w \odot {\tilde Q'},
\end{align}
where the expression for $\tilde Q'$ is shown in \eqref{qref}.
\begin{figure*}[h]
\begin{equation}
\tilde Q'=\left(\mbox{diag}\{(B_i^{\top} P_iP_iB_i)^{-1}B_i^\top P_i\}\right)\tilde{Q}\left(\mbox{diag}\{P_iB_i(B_i^{\top}P_iP_iB_i)^{-1}\}\right). \label{qref}
\end{equation}
\end{figure*}
$\tilde{Q}'$ is close to $\tilde{Q}$ in the least square sense. The drawback of the approximate controller \eqref{eq:K}, therefore, is that instead of minimizing the original objective function \eqref{J}, it minimizes
\begin{equation}
    J = \int_0^\infty \mathbf{x}^\top Q'\mathbf{x} + \mathbf{u}^\top \mathcal{R} \mathbf{u} \,\, dt, \label{app}
\end{equation}
where $Q'=\bar{Q}+L_w \odot \tilde{Q}'$, and $\mathcal R$ follows from \eqref{rtil}.

Also, because $\mathcal P \mathcal B$ and $\mathcal B^\top \mathcal P$ are block diagonal, they only represent different scalings of the agent states in \eqref{ls}. Thus, any communication structure imposed in $\tilde Q$ is preserved in $\tilde R^*$ and $\tilde R^* \mathcal{B}^\top \mathcal P$, which is the global control gain. To implement $\tilde R^*$, neighboring agents need to share their $\mathcal P \mathcal B$ vectors. The loss from the optimal objective function \eqref{J} to the approximated objective function \eqref{app} (or, equivalently the loss from $\tilde{Q}$ to $\tilde{Q'}$) can be numerically shown to become smaller as the number of agents increases. Thus, the true benefit of the design is when the number of agents is large. Theorem 4.1 in \cite{gahinet} can be used to exactly quantify the loss in $J$ in terms of the difference between $\tilde{Q}$ and $\tilde{Q'}$. We skip that derivation, and refer the interested reader to this theorem.

\section{Controller Design using RL}
Equation \eqref{locric} indicates that each local controller can be learned independently using measurements of the local group-level states. The global controller, on the other hand, does not need to be learned owing to the block-diagonal structure of $\mathcal A$ and $\mathcal B$. Once the local controllers are learned, the global controller can be simply computed as the second component on the right hand side of \eqref{eq:K}. Algorithm 1 lists the detailed steps for learning the local RL controllers using ADP based on the approximate LQR in  \eqref{app}. An important point to note is that the group-level state matrix $A_j$ in \eqref{tr} for formulating the LQR problem is assumed to be block-diagonal. However, since ADP is a model-free design, Algorithm 1 is applicable even if $A_j$ is not block-diagonal. We will encounter this scenario in our target-tracking example in Section IV, where we will show that Algorithm 1 still successfully learns the desired model-free controller within a short learning time. From \cite{jiang} it follows that if the exploration noise $\mathbf{u}_{0i}(t)$ is persistently exciting, then  $K^{\text{local}}$ (and $K^{\text{global}}$ computed from it) in Algorithm 1 will asymptotically converge to the respective solutions of the modified LQR problem \eqref{app}.

\begin{algorithm}[H]
\caption{ Off-policy ADP for Hierarchical Controller \eqref{eq:K} }\label{alg:1}
\textbf{Step 1 - Data storage:}
Each group $i=1,\dots, N$ is assigned a coordinator, say denoted as $\mathcal C_i$, that stores $\mathbf{x}_i(t)$ and exploration noise $\mathbf{u}_{0i}(t)$ for an interval $(t_1,t_2,\cdots,t_l)$, with  sampling time $T$. Total data storage time is $T\times$ number of learning time steps. Assume that there exists a sufficiently large number of sampling intervals for each control iteration step such that rank($I_{\mathbf{x}_i} \;\; I_{\mathbf{x}_i}\mathbf{u}_{0i}) = n(n+1)/2 + nm$. This rank condition makes sure that the system is persistently excited. Coordinator $\mathcal C_i$ {constructs} the following matrices:
\begin{align}
&  \delta_{\mathbf{x}_i} = \begin{bmatrix}
\mathbf{x}_i \otimes \mathbf{x}_i |_{t_1}^{t_1+T} ,& \cdots &, \mathbf{x}_i \otimes \mathbf{x}_i |_{t_l}^{t_l+T}
\end{bmatrix}^\top, \nonumber \\
&  I_{\mathbf{x}_i} = \begin{bmatrix}
\int_{t_1}^{t_1+T}(\mathbf{x}_i \otimes \mathbf{x}_i) d\tau ,& \cdots &, \int_{t_l}^{t_l+T} (\mathbf{x}_i \otimes \mathbf{x}_i) d\tau \\
\end{bmatrix} ^\top, \nonumber \\
&  I_{\mathbf{x}_i\mathbf{u}_{0i}} = \begin{bmatrix}
\int_{t_1}^{t_1+T}(\mathbf{x}_i \otimes \mathbf{u}_{0i}) d\tau ,& \cdots & ,\int_{t_l}^{t_l+T} (\mathbf{x}_i \otimes \mathbf{u}_{0i}) d\tau \\
\end{bmatrix} ^\top. \nonumber
\end{align}
\vspace{0.02in}
\textbf{Step 2 - Learning step:}
Starting with a stabilizing controller $K^{\text{local}}_{0i}$, coordinator  $\mathcal C_i$ solves for $P_i$ and $K^{\text{local}}_i$ iteratively as:
\begin{multline}\label{eq:updateA1}
 \underbrace{\begin{bmatrix}
\delta_{\mathbf{x}_i} & -2I_{\mathbf{x}_i}(I_{p_in} \otimes (K^{\text{local}}_{i,k})^\top R_i)  -2I_{\mathbf{x}_i\mathbf{u}_{0i}}(I_{p_in} \otimes R_i)
\end{bmatrix}}_{\Theta_{i,k}}\nonumber \\\begin{bmatrix}
\mbox{vec}(P_{i,k}) \\ \mbox{vec}(K^{\text{local}}_{i,k+1})
\end{bmatrix} =\underbrace{-I_{\mathbf{x}_i}\mbox{vec}(\bar{Q}_{i,k})}_{\Phi_{i,k}}.
\end{multline}
$P_{i,k}$ and $K^{\text{local}}_{i,k+1}$ are iterated till $|P_{i,k} - P_{i,k-1}| < {\epsilon}$, where ${\epsilon}>0$ is a chosen small threshold. 

\vspace{0.06in}

\textbf{Step 3 - Computing global controller :} Once the learning step converges, $\tilde{R}^\ast$ is computed distributively between the coordinators following \eqref{ls}. {Since $K^{\text{local}}_{i,k}$ converges to $R_i^{-1}B_i^TP_i$ and since $R_i$ is known, $B_i^TP_i$ is available.} Thereafter, the coordinator $\mathcal C_i$ computes the global control input for the $i^{th}$ group following \eqref{eq:K} as the $i^{th}$ block of the following vector:
\begin{equation}
    \textbf{u}_g=\tilde{R}^*RK^{\text{local}}\mathbf{x}.
\end{equation}
Considering the $i^{th}$ row of $\tilde{R}^\ast$ is available to $\mathcal C_i$, this implies that the $i^{th}$ co-ordinator must share $R_i K^{\text{local}}_i \textbf{x}_i$ with its {neighboring coordinators}, and vice versa, to compute their respective global controllers distributively.

\vspace{0.06in}

\textbf{Step 4 - Applying joint controller :} Finally, every agent $j$ in the $i^{th}$ group actuates their control signal as
\begin{equation}
    \textbf{u}_{ij}=\{K^{\text{local}}_i \mathbf{x}_i\}(j) + \{\textbf{u}_{g,i}\}(j)
\end{equation}
where, $\textbf{u}_{g,i}$ is the $i^{th}$ block of $\textbf{u}_g$, and $\{\}(j)$ means the $j^{th}$ element of the vector contained in $\{\}$, $i=1,\dots,N$.
\end{algorithm}

\section{Application to Formation control}
We next demonstrate how to make use of the proposed hierarchical learning algorithm for formation control and target tracking applications in multi-agent systems. We first show how this problem can be posed in terms of the optimal control formulation in~\eqref{J}, and then demonstrate the performance of the learning algorithm using a simulation example.

\subsection{Problem formulation}
We consider $p$ robots, whose dynamics are given by
\begin{equation}
    m_i\ddot q_i+c_i\dot q_i = u_i,\quad i = 1,\cdots,p,
\end{equation}
where $q_i\in\mathbb{R}^2$ denotes the 2D position of robot $i$, $m_i\in\mathbb{R}_+$ is the mass of agent $i$, $c_i\in\mathbb{R}_+$ is a damping coefficient that models friction and drag effects, and $u_i\in\mathbb{R}^2$ is the force that acts as a control input.

Denote $x_i = [q_i^\top~\dot q_i^\top]^\top$. We have
\begin{eqnarray}
    \dot x_i&=&G_ix_i+H_iu_i,\label{eq:agent1}\\
    G_i &=& \begin{pmatrix}
    0_2&I_2\\
    0_2&-\frac{c_i}{m_i}I_2
    \end{pmatrix},\quad H_i = \begin{pmatrix}
    0_2\\
    \frac{1}{m_i}I_2
    \end{pmatrix}.\label{eq:agent2}
\end{eqnarray}
We assume that $c_i$ and $m_i$ are unknown parameters.




The robots are divided into $N$ groups to track $N$ different targets. Each group has $p_j$ robots. The state of $i$th agent within group $j$ is denoted by $x_i^j$, $i=1,\cdots, p_j$.  We assume that the locations of the targets, $q^j_T(t)$, $j=1,\cdots,N$, are known and that target assignment is completed so that each group has the knowledge of its assigned target.

The control objective is to ensure that each group converges to a desired formation with its assigned target at the center of the formation while keeping the groups as close as possible, e.g., to maintain a connected communication network. Specifically, for the formation control objective, we choose a reference agent, say agent $1$ in group $j$, and require
\begin{equation}\label{eq:local_obj}
    \left|q^j_i-q^j_{1}-q^{j,d}_{i}\right|\rightarrow 0,~\forall i\in\{2,\cdots,p_j\},
\end{equation}
for some predesigned $q^{j,d}_{i}$.
For the target tracking objective, we require
\begin{equation}
    \left|\sum_{i=1}^{p_j}\frac{1}{N}q^j_i-q_T^j\right|\rightarrow0,\quad \forall j.
\end{equation}
To keep the groups close, we choose to minimize the distance between the centroids of the groups.

We next formulate these  objectives as the optimal control problem \eqref{J}. Towards this end, we rewrite the agent dynamics within a group as
\begin{equation}
    \mathbf{\dot x}_j = A_j\mathbf{x}_j+B_j\mathbf{u}_j
\end{equation}
where $\mathbf{x}_j=\begin{bmatrix} (x^j_1)^\top & \cdots & ( x^j_{p_j})^\top \end{bmatrix}^\top$, $A_j = \mbox{diag}\left\{A^j_1,\cdots,A^j_{p_j}\right\}$, $B_j = \mbox{diag}\left\{B^j_1,\cdots,B^j_{p_j}\right\}$ and $\mathbf{u}_j = \begin{bmatrix} (u^j_1)^\top & \cdots & (u^j_{p_j})^\top \end{bmatrix}^\top$. 

Given the dynamics of $\mathbf{x}_j$, we consider a coordinate transformation $T$ such as
\begin{equation}
    \mathbf{z}_j:=T\mathbf{x}_j = \begin{pmatrix}
    x_2^j-x_1^j\\
    x_3^j-x_1^j\\
    \vdots\\
    x_{p_j}^j-x_1^j\\
    \frac{1}{p_j}\sum_{i=1}^{p_j}{x_i^{j}}
    \end{pmatrix}=\begin{pmatrix}
    \tilde z_1^{j}\\
    \tilde z_2^{j}\\
    \vdots\\
    \tilde z_{p_j-1}^{j}\\
    \bar z^j
    \end{pmatrix}.
\end{equation}
Then the dynamics of $\mathbf{z}_j$ is given by
\begin{equation}\label{eq:z_dot}
    \mathbf{\dot z}_j=T{A}_jT^{-1}\mathbf{z}_j+T{B}_j\mathbf{u}_j.
\end{equation}



Note that $\tilde z^j_i$ includes both relative position and velocity between agent $i+1$ and agent $1$. Let $C=[I_2~0_2]$. Thus, for the formation control objective, we specify desired setpoints of $C\tilde z^j_i$ as $q^{j,d}_{i+1}$,  $i=1,\cdots,p_j-1$. Similarly, for the centroid tracking objective, we specify the setpoint of $C\bar z^j$ as $ q_T^j$.  


Because the setpoints for $C\bar z^j$ and $C\tilde z^j_i$ are non-zero, we take a Linear Quadratic Integral (LQI) control approach~\cite{young} and introduce an integral control to~\eqref{eq:z_dot}. Let $\bar q^j = [(q^{j,d}_{2})^\top,\cdots,(q^{j,d}_{p_j})^\top,(q_T^j)^\top]^\top$. We define the integral control as
\begin{equation}\label{eq:int}
    \mathbf{\dot\zeta}_j=(I_{p_j}\otimes C)\mathbf{z}_j-\bar q^j.
\end{equation}
Let $X_j = [\mathbf{z}_j^\top,\zeta_j^\top]^\top$. The formation control and target tracking objectives for group $j$ can be achieved by minimizing the  objective function
\begin{align}
    J_j&=\int_0^\infty \begin{pmatrix}
    \mathbf{z}_j\\
    \zeta_j
    \end{pmatrix}^\top \bar Q_j\begin{pmatrix}
    \mathbf{z}_j\\
    \zeta_j
    \end{pmatrix} + \mathbf{u}_j^\top R_j\mathbf{u}_j\,dt \nonumber \\ &=\int_0^\infty X_j^\top\bar Q_jX_j + \mathbf{u}_j^\top R_j\mathbf{u}_j\,dt,\label{eq:Jj}
\end{align}
where \begin{equation}
    \bar Q_j = \begin{pmatrix}
    \bar Q_{z,j}&\bar Q_{z\zeta,j}\\
    \bar Q_{\zeta z,j}&\bar Q_{\zeta,j}
    \end{pmatrix}\geq 0,\quad \bar Q_{\zeta,j}>0.
\end{equation}
Because the LQI control minimizing~\eqref{eq:Jj} stabilizes the closed-loop system, we guarantee $\dot\zeta_j\rightarrow0$, which means $(I_{p_j}\otimes C)\mathbf{z}_j\rightarrow\bar q^j$ for constant $\bar q^j$. The stabilizing LQI gains can be learned using Algorithm 1 without $\bar q^j$.

We define $X =[X^\top_1,\cdots,X^\top_N]^\top$ and let $S$ be a matrix such that $\bar z^j = SX_j$. We further define
\begin{equation}
    \bar z = [(\bar z^1)^\top,\cdots,(\bar z^N)^\top]^\top
\end{equation}
which consists of the centroids of all the groups. Note that $\bar z = (I_N\otimes S)X$ and $\bar z^\top(L_w\otimes I_n)\bar z=X^\top(L_w\otimes S^\top S)X$.

To minimize inter-group distance given a communication topology $L_w$, we define the global objective function
\begin{equation}
    J_g=\int_0^\infty X^\top(L_w\otimes S^\top S)X\,dt.
\end{equation}
Optimizing $J_g$ will constrain the motion of the centroids to be close to their neighbors. Let $\tilde Q=(L_w\otimes S^\top S)$, $\mathbf{u} = [\mathbf{u}_1^\top,\cdots,\mathbf{u}_N^\top]^\top$, $\bar Q = \mbox{diag}\{\bar Q_1,\cdots,\bar Q_N\}$, and $R = \mbox{diag}\{R_1,\cdots,R_N\}$. The overall objective function is given by
\begin{equation}
    J = \sum_{j=1}^NJ_j + J_g=\int_0^\infty X^\top(\bar Q+\tilde Q)X + \mathbf{u}^\top R\mathbf{u}\,dt
\end{equation}
which is in the form of~\eqref{J}.
\subsection{Simulation example}
We consider 4 groups of robots with group 1 and 4 having 3 agents and group 2 and 3 having 4 agents, i.e., $p_1=p_4=3$, $p_2=p_3=4$. We assume that the dynamics of the agents in~\eqref{eq:agent1}--\eqref{eq:agent2} are the same within each group. The 4 targets are located at $[5,5],~[5,-5],~[-5,5],[-5,-5]$ meters. The initial conditions of the agents are randomly generated. The communication topology between the groups is a star graph where group $1$, $2$, and $3$ have bidirectional communication with group $4$. We set $\bar Q_j=0.1I$, $R_j = I$, and $\tilde Q = 0.1(L\otimes S^TS)$, where $L$ is the unweighted graph Laplacian matrix. The mass $m_j$ and damping $c_j$ for group $j$ are set to $j$ and $0.1/j$, respectively.

To learn the controller using Algorithm~\ref{alg:1}, we add exploration noise for the initial 6, 15, 15, 6 seconds for the four groups, respectively. The sampling time $T$ of data during the initial learning period is 0.01 seconds. The desired formation of each group is an equilateral triangle and a square for the 3-agent groups and for the 4-agent groups, respectively. The side length of each polygon is $1$.

\begin{figure}[htpb]
    \centering
    \psfrag{y (meters)}{\small{$y$ (meters)}}
    \psfrag{x (meters)}{\small{$x$ (meters)}}
    \psfrag{Trajectories of the agents: solid lines -- optimal control; dash-dot lines: learned approximate control}{\small{}}
    \includegraphics[trim={4cm 0 6.5cm 0},clip,width=0.8\textwidth]{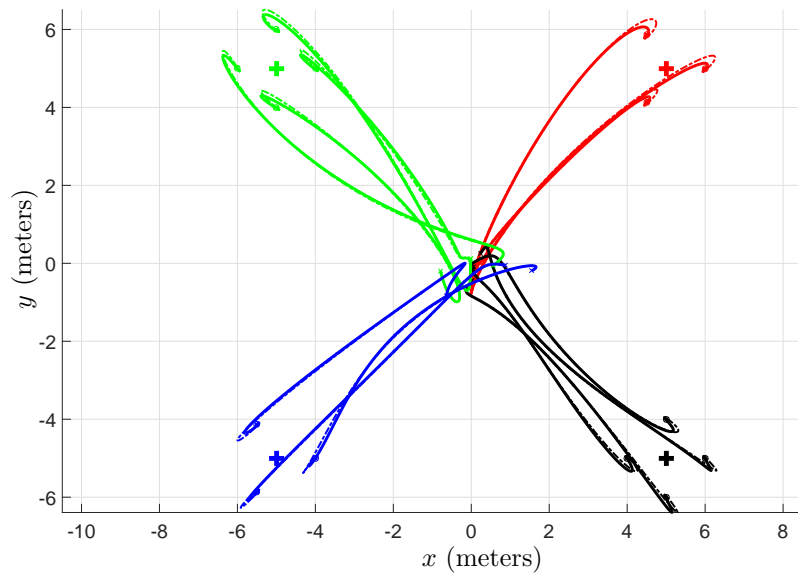}
\caption{Trajectories of the agents for $\tilde{Q} = 0.1\left(L\otimes S^\top S\right)$. Solid line: optimal control. Dash-dotted line: learned approximate control. Targets are denoted by $\bm{+}$'s. Red, black, green, and blue colors indicate group 1 to 4, respectively. }\label{fig:traj1}
\end{figure}
\begin{figure}[htpb]
    \centering
    \psfrag{y (meters)}{\small{$y$ (meters)}}
    \psfrag{x (meters)}{\small{$x$ (meters)}}
    \psfrag{Trajectories of the agents: solid lines -- optimal control; dash-dot lines: learned approximate control}{\small{}}
    \includegraphics[trim={4cm 0 6.5cm 0},clip,width=0.8\textwidth]{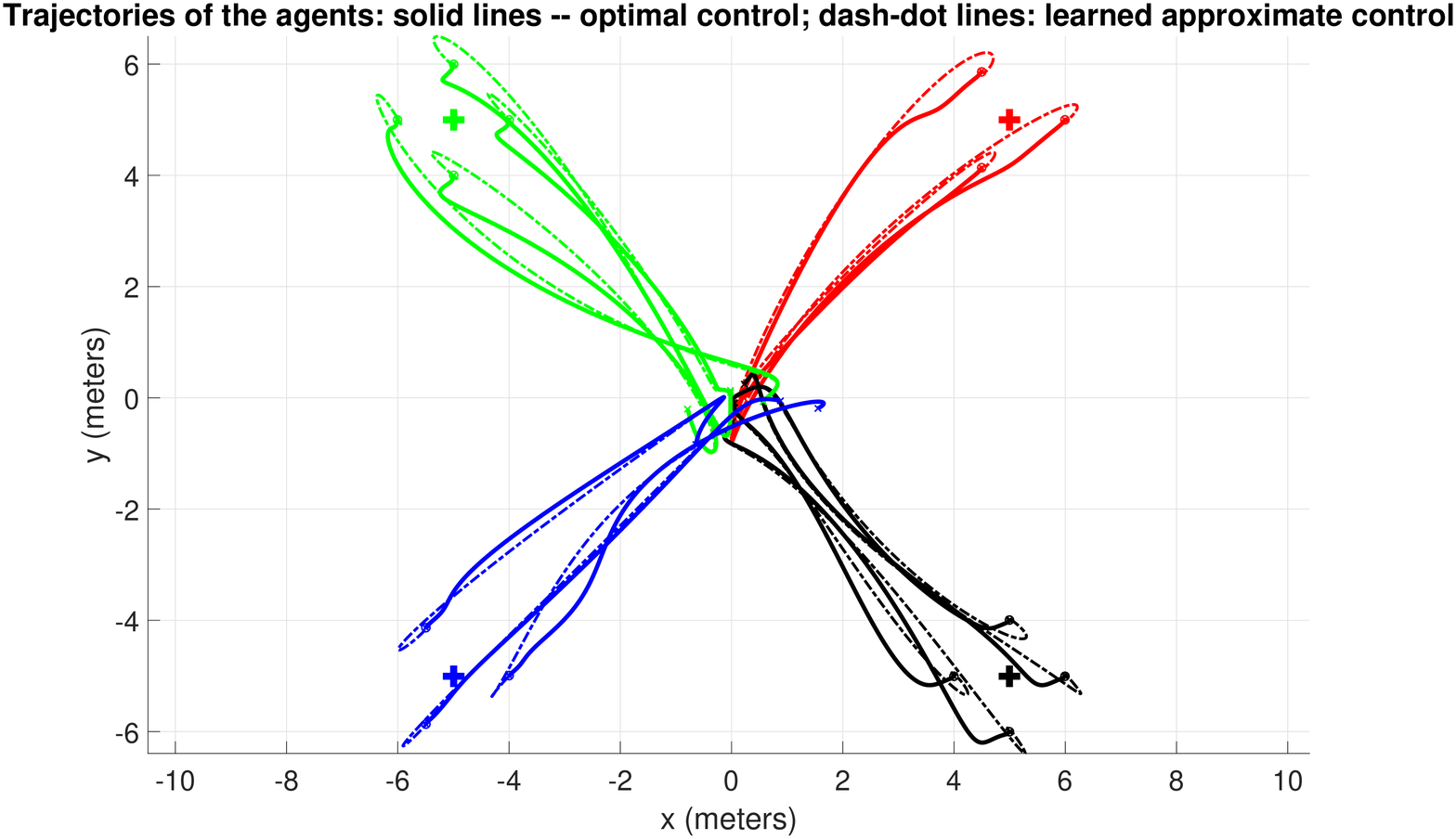}
\caption{Trajectories of the agents for $\tilde{Q} = \left(L\otimes S^\top S\right)$. Solid line: optimal control. Dash-dotted line: learned approximate control. Targets are denoted by $\bm{+}$'s. Red, black, green, and blue colors indicate group 1 to 4, respectively.}\label{fig:traj2}
\end{figure}
\begin{figure}[t]
    \centering
    \includegraphics[trim={4cm 0 25cm 0},clip,width=0.9\textwidth]{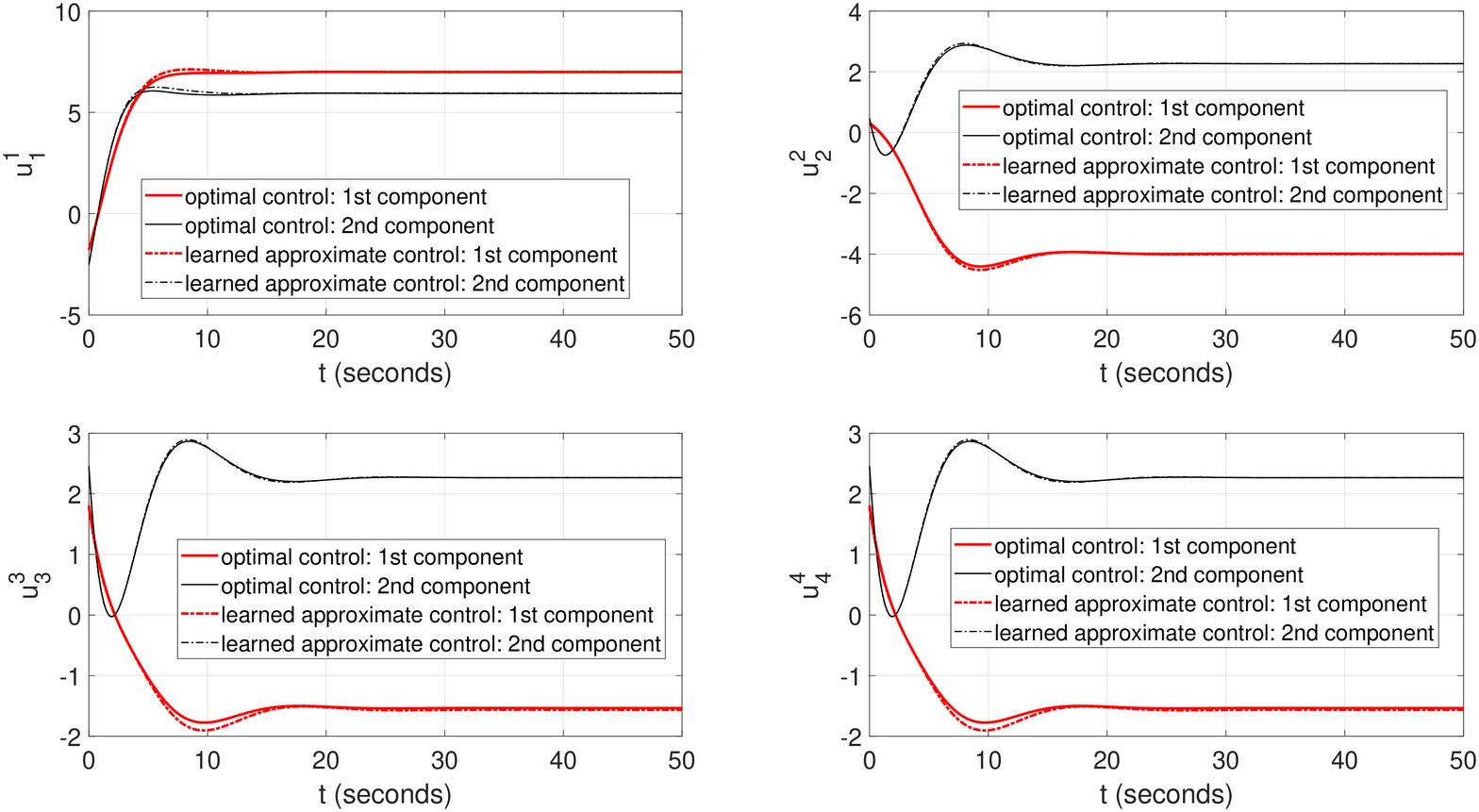}
\caption{Comparison of control inputs for two agents (agent 1 and 3 from group 1 and 3, respectively) between optimal control (solid lines) and learned approximate control (dash-dotted lines). }\label{fig:u1}
\end{figure}
\begin{figure}[htpb]
    \centering
    \includegraphics[trim={4cm 0 25cm 0},clip,width=0.9\textwidth]{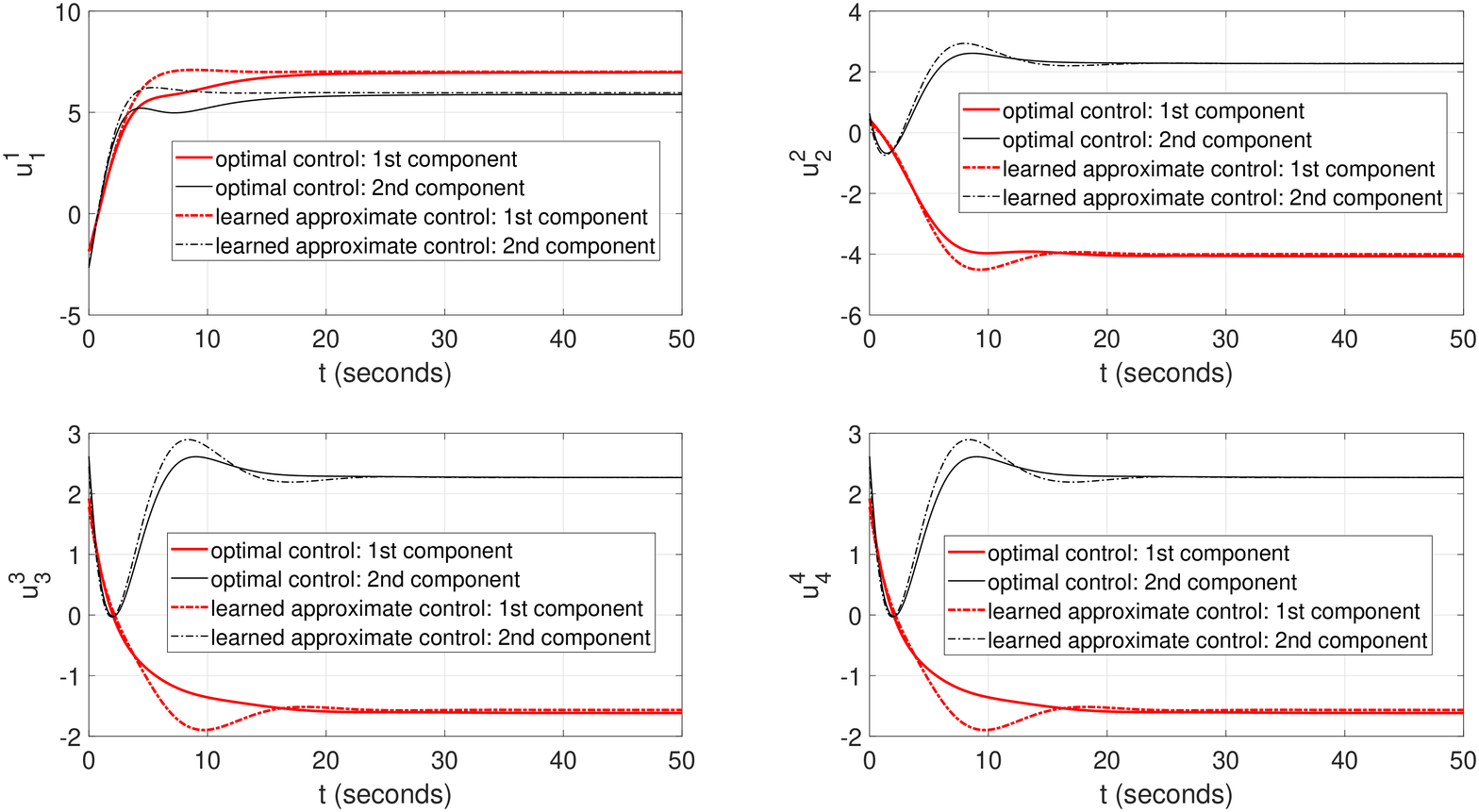}
\caption{Comparison of control inputs for two agents (agent 1 and 3 from group 1 and 3, respectively) between optimal control (solid lines) and learned approximate control (dash-dotted lines) when $\tilde Q$ is increased 10 times.}\label{fig:u2}
\end{figure}

 After learning, the control gains from Algorithm~\ref{alg:1} are implemented. Figure~\ref{fig:traj1} shows the comparison between the trajectories generated from the optimal control and the learned approximate control. As one can see, the learned approximate control achieves the formation control and target tracking objectives. It also yields similar agent trajectories to the optimal control. In Figure~\ref{fig:u1}, the control inputs are compared for agent 1 and 3 from group 1 and 3, respectively, which shows almost the same performance. Thus, the learned control approximately recovers the optimal control performance. As $\tilde Q$ increases, the discrepancy between the two controls becomes more visible. Figure~\ref{fig:traj2} and~\ref{fig:u2} show the same comparison between the trajectories and the control inputs, respectively, when $\tilde Q$ is increased 10 times. We observe from Fig.~\ref{fig:traj2} that although the learned approximate control achieves the formation control and target tracking objectives, the agent trajectories exhibit observable differences from the true optimal trajectories. Similarly, the difference between the learned control and the optimal control is more pronounced in Fig.~\ref{fig:u2} than in Fig.~\ref{fig:u1}.

\section{Conclusion}
We propose a hierarchical LQR control design using model-free reinforcement learning. The design can address global and local control objectives for large multi-agent systems with unknown heterogeneous LTI dynamics, by dividing the agents into distinct groups. The local control for all groups can be learned in parallel, and the global control can be computed algebraically from it, thereby saving learning time. In our future work, we would like to investigate how the RL loops can be implemented in a distributed way  in case the open-loop dynamics of the agents are coupled. 

\end{document}